\documentclass[a4paper,12pt]{article}

\catcode`\@=11 \long\def\@makefntext#1{ \protect\noindent \hbox to 3.2pt
{\hskip-.9pt
$^{{\eightrm\@thefnmark}}$\hfil}#1\hfill}       

\def\@makefnmark{\hbox to 0pt{$^{\@thefnmark}$\hss}}    

\def\ps@myheadings{\let\@mkboth\@gobbletwo
\def\@oddhead{\hbox{}
\rightmark\hfil\eightrm\thepage}
\def\@oddfoot{}\def\@evenhead{\eightrm\thepage\hfil
\leftmark\hbox{}}\def\@evenfoot{}
\def\sectionmark##1{}\def\subsectionmark##1{}}

\oddsidemargin=\evensidemargin 
\newcounter{sectionc}\newcounter{subsectionc}\newcounter{subsubsectionc}
\renewcommand{\section}[1] {\vspace{12pt}\addtocounter{sectionc}{1}
\setcounter{subsectionc}{0}\setcounter{subsubsectionc}{0}\noindent
    {\tenbf\thesectionc. #1}\par\vspace{5pt}}
\renewcommand{\subsection}[1] {\vspace{12pt}\addtocounter{subsectionc}{1}
    \setcounter{subsubsectionc}{0}\noindent
    {\bf\thesectionc.\thesubsectionc. {\kern1pt \bfit #1}}\par\vspace{5pt}}
\renewcommand{\subsubsection}[1] {\vspace{12pt}\addtocounter{subsubsectionc}{1}
    \noindent{\tenrm\thesectionc.\thesubsectionc.\thesubsubsectionc.
    {\kern1pt \tenit #1}}\par\vspace{5pt}}

\newcounter{appendixc}
\newcounter{subappendixc}[appendixc]
\newcounter{subsubappendixc}[subappendixc]
\renewcommand{\thesubappendixc}{\Alph{appendixc}.\arabic{subappendixc}}
\renewcommand{\thesubsubappendixc}
    {\Alph{appendixc}.\arabic{subappendixc}.\arabic{subsubappendixc}}

\renewcommand{\appendix}[1] {\vspace{12pt}
        \refstepcounter{appendixc}
        \setcounter{figure}{0}
        \setcounter{table}{0}
        \setcounter{lemma}{0}
        \setcounter{theorem}{0}
        \setcounter{corollary}{0}
        \setcounter{definition}{0}
        \setcounter{equation}{0}
        \renewcommand{\thefigure}{\Alph{appendixc}.\arabic{figure}}
        \renewcommand{\thetable}{\Alph{appendixc}.\arabic{table}}
        \renewcommand{\theappendixc}{\Alph{appendixc}}
        \renewcommand{\thelemma}{\Alph{appendixc}.\arabic{lemma}}
        \renewcommand{\thetheorem}{\Alph{appendixc}.\arabic{theorem}}
        \renewcommand{\thedefinition}{\Alph{appendixc}.\arabic{definition}}
        \renewcommand{\thecorollary}{\Alph{appendixc}.\arabic{corollary}}
        \renewcommand{\theequation}{\Alph{appendixc}.\arabic{equation}}
        \noindent{\tenbf Appendix \theappendixc #1}\par\vspace{5pt}}
\newcommand{\subappendix}[1] {\vspace{12pt}
        \refstepcounter{subappendixc}
        \noindent{\bf Appendix \thesubappendixc. {\kern1pt \bfit #1}}
    \par\vspace{5pt}}
\newcommand{\subsubappendix}[1] {\vspace{12pt}
        \refstepcounter{subsubappendixc}
        \noindent{\rm Appendix \thesubsubappendixc. {\kern1pt \tenit #1}}
    \par\vspace{5pt}}

\newcommand{\textlineskip}{\baselineskip=13pt}
\newcommand{\smalllineskip}{\baselineskip=10pt}

\def\eightcirc{
\begin{picture}(0,0)
\put(4.4,1.8){\circle{6.5}}
\end{picture}}
\def\eightcopyright{\eightcirc\kern2.7pt\hbox{\eightrm c}}

\def\abstracts#1#2#3{{
    \centering{\begin{minipage}{4.5in}\footnotesize\baselineskip=10pt
    \parindent=0pt #1\par
    \parindent=15pt #2\par
    \parindent=15pt #3
    \end{minipage}}\par}}

\newcounter{itemlistc}
\newcounter{romanlistc}
\newcounter{alphlistc}
\newcounter{arabiclistc}

\newcommand{\fcaption}[1]{
        \refstepcounter{figure}
        \setbox\@tempboxa = \hbox{\footnotesize Fig.~\thefigure. #1}
        \ifdim \wd\@tempboxa > 5in
           {\begin{center}
        \parbox{5in}{\footnotesize\smalllineskip Fig.~\thefigure. #1}
            \end{center}}
        \else
             {\begin{center}
             {\footnotesize Fig.~\thefigure. #1}
              \end{center}}
        \fi}

\newcommand{\tcaption}[1]{
        \refstepcounter{table}
        \setbox\@tempboxa = \hbox{\footnotesize Table~\thetable. #1}
        \ifdim \wd\@tempboxa > 5in
           {\begin{center}
        \parbox{5in}{\footnotesize\smalllineskip Table~\thetable. #1}
            \end{center}}
        \else
             {\begin{center}
             {\footnotesize Table~\thetable. #1}
              \end{center}}
        \fi}

\def\@citex[#1]#2{\if@filesw\immediate\write\@auxout
    {\string\citation{#2}}\fi
\def\@citea{}\@cite{\@for\@citeb:=#2\do
    {\@citea\def\@citea{,}\@ifundefined
    {b@\@citeb}{{\bf ?}\@warning
    {Citation `\@citeb' on page \thepage \space undefined}}
    {\csname b@\@citeb\endcsname}}}{#1}}

\newif\if@cghi
\def\cite{\@cghitrue\@ifnextchar [{\@tempswatrue
    \@citex}{\@tempswafalse\@citex[]}}
\def\citelow{\@cghifalse\@ifnextchar [{\@tempswatrue
    \@citex}{\@tempswafalse\@citex[]}}
\def\@cite#1#2{{$\null^{#1}$\if@tempswa\typeout
    {IJCGA warning: optional citation argument
    ignored: `#2'} \fi}}

\def\pmb#1{\setbox0=\hbox{#1}
    \kern-.025em\copy0\kern-\wd0
    \kern.05em\copy0\kern-\wd0
    \kern-.025em\raise.0433em\box0}

\def\fnt#1#2{\footnotetext{\kern-.3em
    {$^{\mbox{\scriptsize #1}}$}{#2}}}

\def\fpage#1{\begingroup
\voffset=.3in \thispagestyle{empty}\begin{table}[b]\centerline{\footnotesize #1}
    \end{table}\endgroup}

\def\runninghead#1#2{\pagestyle{myheadings}
\markboth{{\protect\footnotesize\it{\quad #1}}\hfill}
{\hfill{\protect\footnotesize\it{#2\quad}}}} \headsep=15pt

\font\tenrm=cmr10 \font\tenit=cmti10 \font\tenbf=cmbx10 \font\bfit=cmbxti10 at 10pt
\font\ninerm=cmr9   \font\eightrm=cmr8

\def\qed{\hbox{${\vcenter{\vbox{            
   \hrule height 0.4pt\hbox{\vrule width 0.4pt height 6pt
   \kern5pt\vrule width 0.4pt}\hrule height 0.4pt}}}$}}

\begin{document}
\setlength{\textheight}{7.7truein}  

\runninghead{Gravitational Constant and Torsion}{Gravitational Constant and Torsion
}

\normalsize\textlineskip \thispagestyle{empty} \setcounter{page}{1}



\fpage{1} \centerline{\bf  GRAVITATIONAL CONSTANT AND TORSION}
\centerline{\footnotesize PRASANTA MAHATO\footnote{e-mail:
pmahato@dataone.in}}
\centerline{\footnotesize\it Department of Mathematics, Narasinha Dutt College}
\baselineskip=10pt \centerline{\footnotesize\it HOWRAH,West Bengal,INDIA 711101}
\vspace*{10pt}

\vspace*{0.225truein}




 \abstracts{Riemann-Cartan space time $U_{4}$ is considered here. It has  been shown that when we link
topological Nieh-Yan density with the gravitational constant then we get
Einstein-Hilbert Lagrangian as a consequence.}{PACS numbers : 04.20.Cv,
04.20.Fy}{Key words : Pontryagin density, Torsion, Gravitational constant,
Einstein-Hilbert Lagrangin}
 \section{Introduction}

 A generic Riemann manifold is endowed with two fundamental and independent
 entities: a metric and an affine connection. If the affine connection is not
 assumed to be a function of the metric, then the local geometry is endowed with two
 independent tensors, curvature and torsion. It was Einstein's point of view - that
 torsion would be an unnecessary addition which was not required for the most
 economical and successful theory of space time. Cartan on the other hand refused to
 accept it on the ground that the two notions are logically independent and
 therefore Einstein's proposal was a particular case. For more details on this
 controversy one can see the letters exchanged between them between 1929 and 1932\cite{Deb79}.
 In recent years, much more abstract geometrical frameworks are modelled to handle
 many pertinent issues of particle physics especially for a consistent quantum
 theory of gravitation. In this setup torsion does not look a strange idea.

 Curvature plays an important role in the characterization of the topological
 structure of the manifold. It is a remarkable result of differential geometry that
 certain global features of a manifold are determined by some local functionals of
 its intrinsic geometry. Pontryagin and Euler classes are two well known examples in
 four dimension. Values of these invariants depend on the global properties of the
 manifold. These invariants are expected to be related to the global values of some
 physical observables.

 The topological features associated with the gauge orbit space of a non-Abelian
 gauge theory when the topological $\theta$-term is introduced in the Lagrangian
 corresponds to a vortex line, and the gauge orbit space appears to be multiply
 connected in nature. This has an implication in the loop space formalism, in the
 sense that the latter involves nonlocality and there is no way we could arrive at a
 corresponding continuum limit. In the gravity with out the metric formalism, it has
 been observed that the $\theta$-term in the Lagrangian effectively corresponds to
 the introduction of torsion and it is found to be associated with the vortex line\cite{Ban96} and
the cosmological constant\cite{Mah02} .

In the following section we shall review some topological aspects of Riemann-Cartan
space time. In section 3 we shall try to derive the metric considering the
connection as the primary object and then,  after  interpreting the gravitational
constant as connected to  a diffeomorphism  invariant density of the
 manifold, will get the Einstein-Hilbert Lagrangian
  as a consequence. Section 4 is the section
of discussion.

\section{Axial vector torsion and topological invariants}
 Cartan's structural equations for a Riemann-Cartan space-time $U_{4}$ are given by
 \cite{Car22,Car24,Kob63}
 \begin{eqnarray}T^{a}&=& de^{a}+\omega^{a}{}_{b}\wedge e^{b}\label{eqn:ab}\\
 R^{a}{}_{b}&=&d\omega^{a}{}_{b}+\omega^{a}{}_{c}\wedge \omega^{c}{}_{b},\label{eqn:ac}\end{eqnarray}
here $\omega^{a}{}_{b}$ and e$^{a}$ represent the spin connection
and the local frames respectively.

From these two expressions, curvature seems to be more fundamental than torsion.
Definition (2) requires only connection whereas definition (1) requires both
connection and local frame. But, since on any smooth  metric manifold a local frame
is necessarily always defined, torsion can exist even if the connection vanishes.
This implies torsion and curvature should be treated on a similar footing. Torsion
appears rather naturally in the commutator of two covariant derivatives for the
group of diffeomorphisms of a manifold in a coordinate basis\cite{Cho82,Lov89},
\begin{eqnarray} [\nabla_{\mu},\nabla_{\nu}]V^{A}&=&-T^{\lambda}{}_{\mu\nu}\nabla_{\lambda}V^{A}
+R^{A}{}_{B\mu\nu}V^{B}
\end{eqnarray}
where $V^{A}$ represents any tensor (or spinor) under
diffeomorphisms (or under the group of tangent rotations), and
$R^{A}{}_{B}$ is the curvature tensor in the corresponding
representation. Here curvature and torsion play quite different
roles: $T^{\lambda}{}_{\mu\nu}$ is the structure function for the
diffeomorphism group and $R^{A}{}_{B\mu\nu}$ is central charge.
This indicates that the gauge approach of gravity can be achieved
through the local version of the Lorentz-Poincar\'{e} symmetry
where $R^{A}{}_{B\mu\nu}$   and $T^{\lambda}{}_{\mu\nu}$,
respectively, represent rotational symmetry and translational
symmetry in the tangent space\cite{Kib61,Sci64,Heh76,Dre82}.

In $U_{4}$ there exists  two invariant four forms. One is the well known
Pontryagin\cite{Che74,Che71} density \textit{P} and the other is the less known
Nieh-Yan\cite{Nie82} density \textit{N} given by
\begin{eqnarray} \textit{P}&=& R^{ab}\wedge R_{ab}\\  \mbox{and} \hspace{2 mm}  \textit{N}&=& T^{a}\wedge
T_{a}- R_{ab}\wedge e^{a}\wedge e^{b}.\end{eqnarray} $P$ and $N$ being four forms, $
 \mathcal{P}=\int Pd^{4}x$ and $  \mathcal{N}=\int Nd^{4}x$  are diffeomorphism
invariant quantities representing certain global features of the manifold.
$\mathcal{P}$ is dimensionless whereas $\mathcal{N}$ has the dimension of L$^{2}$.
This naturally suggests a fundamental (length)$^{2}$ constant like Newton's
gravitational constant G or the inverse cosmological constant $\Lambda^{-1}$.
Chandia and Zanelli\cite{Cha97} have shown that if we  combine the spin connection
and the vierbein together in a connection for $SO(5)$, in the tangent space, in the
form
\begin{eqnarray}W^{AB}&=&\left
[\begin{array}{cc}\omega^{ab}&\frac{1}{l}e^{a}\\-\frac{1}{l}e^{b}&0\end{array}\right],\end{eqnarray}
where $a,b = 1,2,..4;A,B = 1,2,..5$ and $l$ is a fundamental length constant, then
we obtain the $SO(5)$ curvature 2-form,
\begin{eqnarray}F^{AB}&=&dW^{AB}
+W^{AC}\wedge W^{CB}\\&=&\left[\begin{array}{cc}R^{ab}-\frac{1}{l^{2}}e^{a}\wedge e^{b}&\frac{1}{l}T^{a}\\
-\frac{1}{l}T^{b}&0\end{array}\right],\end{eqnarray} and the $SO(5)$ Pontryagin
density,
\begin{eqnarray}F^{AB}\wedge F_{AB}&=&R^{ab}\wedge R_{ab}+\frac{2}{l^{2}}[T^{a}\wedge T_{a}-R^{ab}\wedge e_{a}\wedge e_{b}].\label{eqn:xy}\end{eqnarray}
The first term of the right hand side of (\ref{eqn:xy}) is the $SO(4)$ Pontryagin
density and hence we can write,
\begin{eqnarray} \frac{2}{l^{2}}\int_{M_{4}}N&=&
P_{4}[SO(5)]-P_{4}[SO(4)]\\&=& const.\times\frac{l^{2}}{2}(z_{1}+z_{2}+z_{3}),
 \hspace{2 mm} z_{i}\varepsilon Z.\end{eqnarray} Here $P_{4}[G]$ is the Pontryagin class for a
compact group $G$ on a four dimensional manifold.

From (\ref{eqn:ab}) and  (\ref{eqn:ac}) we see that, when the
vierbein is well defined, \begin{eqnarray} N&=& d(e^{a} \wedge
T_{a})\label{eqn:ad}\\&=& T^{a}\wedge T_{a}-e^{a}\wedge
DT_{a}\label{eqn:ae}\\\mbox{where} \hspace{2 mm}DT_{a}&=&
dT_{a}+\omega^{b}{}_{a}\wedge T_{b}\\&=&R_{ab}\wedge
e^{b}.\end{eqnarray} From (\ref{eqn:ad}) we see that the Nieh-Yan
density is the total derivative of a Chern-Simon like term,
\begin{eqnarray}e^{a} \wedge
T_{a}&\sim&\epsilon^{\mu\nu\alpha\beta}T_{\nu\alpha\beta}\\&=&S^{\mu}\end{eqnarray}where
$S^{\mu}$,  the axial vector part of the torsion, only contributes to the three
form.

We can decompose the torsion tensor $T_{\alpha\beta\gamma}$ into its three
irreducible components, given by,
\begin{eqnarray}T_{\beta}&=&T^{\alpha}{}_{\beta\alpha}\\S^{\mu}&=&\epsilon^{\mu\nu\alpha\beta}T_{\nu\alpha\beta}\\\mbox{and}\hspace{2 mm}
q_{\alpha\beta\mu}&=&T_{\alpha\beta\mu}-\frac{1}{3}(T_{\beta}g_{\alpha\mu}-T_{\mu}g_{\alpha\beta})+\frac{1}{6}
\epsilon_{\alpha\beta\mu\nu}S^{\nu}\end{eqnarray} The minimal action of a spin
$\frac{1}{2}$ field $\psi$ with an external gravitational field with torsion is
given by\cite{Sha01},\begin{eqnarray}S_{\frac{1}{2},min}&=&i\int
d^{4}x\sqrt{g}\bar{\psi}(\gamma^{\alpha}
\nabla_{\alpha}-\frac{i}{8}\gamma^{5}\gamma^{\alpha}
S_{\alpha}-im)\psi,\end{eqnarray}where the covariant derivative $ \nabla_{\alpha}$
is torsionless. Here we see that only the axial vector part of the torsion interacts
with the spin $\frac{1}{2}$ field and the other irreducible components of the
torsion tensor, $T_{\alpha}$ and $q_{\alpha\beta\gamma}$, completely decouple from
the action. Therefore having no source in the matter lagrangian we can simply assume
$T_{\alpha}=0$ and $q_{\alpha\beta\gamma}=0$. With this assumption the Nieh-Yan
density  (\ref{eqn:ae}) becomes  \begin{eqnarray}N&=& -e^{a}\wedge
DT_{a}.\end{eqnarray}

In terms of vierbeins we can write the spin connection in $U_{4}$ as\cite{Sha01},
\begin{eqnarray}\omega_{\mu ab}&=&\bar{\omega}_{\mu ab}+\frac{1}{4}K^{\alpha}{}_{\lambda\mu}(e^{\lambda}{}_{a}e_{b\alpha}-e^{\lambda}{}_{b}e_{a\alpha}),\\\mbox{where}\hspace{2
mm}\bar{\omega}_{\mu
 ab}&=&\frac{1}{4}(e_{b\alpha}\partial_{\mu}e^{\alpha}{}_{a}-e_{a\alpha}\partial_{\mu}e^{\alpha}{}_{b})
+\frac{1}{4}\Gamma^{\alpha}{}_{\lambda\mu}(e^{\lambda}{}_{a}e_{b\alpha}-e^{\lambda}{}_{b}e_{a\alpha}).\end{eqnarray}
Here $\Gamma^{\alpha}{}_{\lambda\mu}$ is the ordinary Christoffel
symbol and
$K^{\alpha}{}_{\lambda\mu}=\frac{1}{2}(T^{\alpha}{}_{\lambda\mu}-K_{\lambda}{}^{\alpha}{}_{\mu}-K_{\lambda\mu}{}^{\alpha})$
is the contorsion tensor. By $\nabla_{\alpha}$ here we mean the
torsionless covariant derivative with respect to the connection
$\bar{\omega}_{\mu ab}$ whereas the full-connection $\omega_{\mu
ab}$ gives the $U_{4}$  covariant derivative $D_{\alpha}.$ With
this nomenclature we see that, from (\ref{eqn:ae}), the Nieh-Yan
density takes the following form,\begin{eqnarray}N&=& -e^{a}\wedge DT_{a}\label{eqn:af}\\&\sim&\nabla_{\mu}S^{\mu} \\
&=&\partial_{\mu}S^{\mu}.\end{eqnarray} From (\ref{eqn:af}) it is clear that $N$ is
an invariant under Lorentz rotation in tangent space but it is not in the case of
(A)dS boost there i.e. when $\frac{1}{l}e^{a}$ itself transforms as a gauge field.
There is a known lemma\cite{Zan00} which states that:

\textbf{\emph{Lemma:}} For $d=4k$, the only parity-odd d-forms built from
$e^{a},R^{ab}$ and $T^{a}$, invariant under AdS group, are the Chern characters for
$SO(d+1)$.

This lemma is equally valid in case of Lorentz group $SO(d-1,1)$ when the (A)dS
group is either $SO(d,1),SO(d-1,2)$ or $SO(d+1)$, because the analysis that follows
is insensitive of the signature. For $d=4$ there is only one such AdS invariant, the
second Chern character of the AdS group,  given by\cite{Zan00}
\begin{eqnarray}
 R^{A}{}_{B}\wedge R^{B}{}_{A}&=& R^{a}{}_{b}\wedge R^{b}{}_{a}+\frac{2}{l^{2}}(T^{a}\wedge T_{a}-R^{ab}\wedge
 e_{a}\wedge e _{b}) \label{eqn:ax}.\end{eqnarray}Hence none of $P$ and $N$
 is AdS invariant only their combination
 \begin{eqnarray}P_{4}(SO(5-i,i))_{i=1\hspace{1 mm}\mbox{or}\hspace{1 mm}2}&=&
 P+\frac{2}{l^{2}}N,\end{eqnarray} is (A)dS invariant. And in
 particular in the case of axial vector torsion, i.e. when $T^{a}\wedge T_{a}=0$ in (\ref{eqn:ax}), this invariant has no explicit dependence on torsion.

 \section{Connection, Curvature, Metric and Gravitational Constant}

From the definition of curvature in (\ref{eqn:ac}) we see that
only the knowledge of connection is required. The role of local
frames are only implicit here. From (\ref{eqn:ax}) we see that, in
the case of axial vector torsion,  the $SO(4,1)$ Pontryagin
density reduces to $SO(3,1)$ Pontryagin density unless the torsion
contributing density $R^{ab}\wedge e_{a}\wedge e_{b}\neq 0$. Here
we heuristically try to define vierbeins $e_{\mu}{}^{a}$ from
$SO(3,1)$ curvatures $R_{\mu\nu ab}$ to get the curvature
$R_{\mu\nu\alpha\beta}$ having only external indices. To include
torsion here we forgo  just one property of Einstein's curvature
tensor, viz, $\epsilon^{\alpha\beta\gamma\delta}
R_{\alpha\beta\gamma\mu}=0$. So we define vierbeins
$e_{\mu}{}^{a}$ and the external curvature $R_{\mu\nu\alpha\beta}$
simultaneously given by
\begin{eqnarray} R_{\mu\nu\alpha\beta}&=&R_{\mu\nu a b}e_{\alpha}{}^{a}e_{\beta}{}^{b},\\
\mbox{such that} \hspace{10
mm}R_{\mu\nu\alpha\beta}&=&R_{\alpha\beta\mu\nu}.\label{eqn:xa}\end{eqnarray}
Equation (\ref{eqn:xa}) gives fifteen equations to determine 16 vierbeins, hence
another constraint is required to completely specify the vierbeins.  The natural
answer comes from the Nieh-Yan invariant, if we impose the
condition,\begin{eqnarray} R_{\mu\nu
 ab}e_{\alpha}{}^{a}e_{\beta}{}^{b}\epsilon^{\mu\nu\alpha\beta}=e\acute{g},\label{eqn:xaa}\end{eqnarray}
where $e=det(e_{\mu}{}^{a})$ and $\acute{g}$ is a scalar
(constant!)  of dimension $L^{-2}$. Therefore in one hand we have
36 $R_{\mu\nu a b}$ and 1 $\acute{g}$, i.e. altogether 37
independent components. On the other hand we have 16 vierbeins
$e_{\mu}{}^{a}$ and 21 external curvature components
$R_{\mu\nu\alpha\beta}$, i.e. altogether 37 components. Hence
vierbeins and external curvatures are completely specified in
terms of  the internal $SO(3,1)$ curvatures and one scalar
$\acute{g}$
 representing  torsion. Here, it is to be noted that,
 (\ref{eqn:xa}) together with (\ref{eqn:xaa})  give minimal
 extension to the  Einstein's theory to incorporate torsion.
 Both the constraints can be derived from an
 action principle using standard techniques of Lagrange
 multipliers   such that the
 vierbeins can be varied independently and then the Lagrange
 multipliers can be set to take zero values. Also when we treat the vierbeins as independent
 fields  (\ref{eqn:xa}) and (\ref{eqn:xaa}) impart constraint on the torsion part of the $SO(3,1)$
  connection. Thus degrees of freedom
 of this theory is greater than that of Einstein's theory with
 torsion contributing to the additional degree.

 With respect to the completely antisymmetric tensor $\epsilon_{\mu\nu\alpha\beta}$ we
 can decompose $R_{\mu\nu\alpha\beta}$ into two independent parts, given by,
 \begin{eqnarray}R_{\mu\nu\alpha\beta}&=&\bar{R}_{\mu\nu\alpha\beta}+
 \frac{1}{24}e\acute{g}\epsilon_{\mu\nu\alpha\beta},\\\mbox{such that}\hspace{2 mm}
 \bar{R}_{\mu\nu\alpha\beta}\epsilon^{\mu\nu\alpha\gamma}&=&0\\\mbox{and}\hspace{2 mm}
 R_{\mu\nu\alpha\beta}\epsilon^{\mu\nu\alpha\gamma}&=&\frac{1}{4}e\acute{g} \delta^{\gamma}_{\beta}.\end{eqnarray}
After this decomposition the $SO(3,1)$ Pontryagin density can be written as
\begin{eqnarray}\epsilon^{\mu\nu\alpha\beta}R_{\mu\nu ab}R_{\alpha\beta}{}^{ab}&=&
\epsilon^{\mu\nu\alpha\beta}\bar{R}_{\mu\nu\gamma\delta}\bar{R}_{\alpha\beta}{}^{\gamma\delta}+\frac{1}{6}e\acute{g}R\\
\mbox{where}\hspace{2
mm}R&=&\bar{R}_{\alpha\beta}{}^{\alpha\beta}\\&=&R_{\alpha\beta}{}^{\alpha\beta}.\end{eqnarray}
Here, as  vierbeins are already introduced, use of metric has been
done for raising and lowering the external indices. As  long as
  (\ref{eqn:xa}) and (\ref{eqn:xaa}) are valid,
$SO(4,1)$ Pontryagin density can be written as
\begin{eqnarray}R^{A}{}_{B}\wedge R^{B}{}_{A}&\sim
&\epsilon^{\mu\nu\alpha\beta}\bar{R}_{\mu\nu\gamma\delta}\bar{R}_{\alpha\beta}{}^{\gamma\delta}+\frac{1}{6}e\acute{g}R-
\frac{2}{l^{2}}e\acute{g}.\label{eqn:xb}\end{eqnarray}

It is well known that, if we treat any Pontryagin density as a Lagrangian, it will
contribute nothing locally, at least in the classical level,  but  globally it
signifies some global property of the   gauge field and the manifold. Hence to have
an effective field theory  we can consider either of the first two terms of
(\ref{eqn:xb}), but not both, to produce locally nontrivial action. Therefore we
heuristically propose the gravitational Lagrangian as,
\begin{eqnarray}\mathcal{L}_{G}&=&\epsilon^{\mu\nu\alpha\beta}\bar{R}_{\mu\nu\gamma\delta}\bar{R}_{\alpha\beta}{}^{\gamma\delta}+(a+\frac{1}{6})e\acute{g}R-
\frac{2}{l^{2}}e\acute{g}\hspace{1 mm},\end{eqnarray}where $a$ is a pure number. For
$a=0$ this Lagrangian is locally trivial.  In particular taking $a=1$, locally this
Lagrangian is equivalent to the Einstein-Hilbert
 Lagrangian,\begin{eqnarray}\mathcal{L_{EH}}&=&\frac{1}{\kappa}eR,\end{eqnarray} with $\acute{g}=\frac{1}{\kappa},$ here $\kappa$ is Einstein's gravitational
constant and then Newton's constant of gravitation is given by,\begin{eqnarray}
G&=&\frac{c^{2}}{8\pi \acute{g}}\hspace{4mm}.\label{eqn:xc}\end{eqnarray}

From (\ref{eqn:xc}) there is a possibility that $G$ is a variable when $\acute{g}$,
  the torsional part of the curvature, is a variable. Following Dirac's\cite{Dir38}
  large number hypothesis one can argue\cite{Chi01,Chi99}  that, strictly
speaking, $G$ is not a constant like the  fine structure constant
$\alpha= {\textsl{e}}^{2}$.

 \section{Discussion}

 One of the most interesting problems of elementary particle physics is to
 understand the gravitation or  the gravitational constant. According to Brans-Dicke
  theory, the value of $G$ is determined by the value of the Brans-Dicke scalar field
  $\phi$. The Brans-Dicke version of Einstein-Cartan theory, with nonzero torsion
  and vanishing non-metricity, was discussed by many  authors\cite{Rau84,Ger85,Kim86}.
   In these approaches $\phi$ acts as a source of torsion\cite{Ber93}.  In
our approach $\acute{g}$ has topological origin where in one hand $e\acute{g}$ is
from the torsional curvature of $SO(3,1)$ gauge group and on the other hand it is
the Nieh-Yan density. In a recent paper\cite{Lan01} it has been shown that torsion
is a natural consequence in a non-commutative $U(1)$ Yang-Mills theory where gauge
symmetries give very natural and explicit realizations of the mixing of spacetime
and internal symmetries. Here torsion measures   the noncommutativity of
displacement of points in the flat spacetime in the teleparallel theory and the
noncommutativity scale is given by the Planck length. So this approach is very much
akin to our present approach if we consider that torsion is connected to internal
space which makes the spacetime noncommutative such that the gravitational constant
fixes the scale of noncommutativity. This supports the findings of some other works,
when the gauge  group is $SL(2,C)$ which is the covering group of $SO(3,1)$, where
torsion has been shown to be  linked with CP-violation\cite{Ban95} and fermion
mass\cite{Ban00}.
 \vspace{4 mm}

            \noindent\textbf{Acknowledgement}

              \vspace{1 mm}
           I wish to thank Prof. Pratul Bandyopadhyay, Indian Statistical Institute, for his remarks and suggestions on this
           problem.

 \vspace{4 mm}

            \noindent\textbf{References}

             \bibliographystyle{unsrt}
                \bibliography{tbib}
\end{document}